\documentclass[twocolumn,showpacs,prl]{revtex4}

\usepackage{amsmath,amssymb,graphicx}
\usepackage{color}

\renewcommand{\vec}[1]{\mathbf{#1}}

\newcommand{\nag}{{\phantom{\dag}}}
\newcommand{\oh}{\mbox{$\frac{1}{2}$}}

\begin{document}

\title{Correlation Effects in Quantum Spin-Hall Insulators: A Quantum Monte
  Carlo Study}

\author{M. Hohenadler, T. C. Lang and F. F. Assaad}

\affiliation{\mbox{Institut f\"ur Theoretische Physik und Astrophysik,
    Universit\"at W\"urzburg, Am Hubland, 97074 W\"urzburg, Germany}}

\begin{abstract}
  We consider the Kane-Mele model supplemented by a Hubbard $U$ term. The
  phase diagram is mapped out using projective auxiliary field quantum Monte
  Carlo simulations.  The quantum spin liquid of the Hubbard model is robust
  against weak spin-orbit interaction, and is not adiabatically connected to
  the spin-Hall insulating state. Beyond a critical value of ${U >
    U_\text{c}}$ both states are unstable toward magnetic ordering. In the
  quantum spin-Hall state we study the spin, charge and single-particle
  dynamics of the helical Luttinger liquid by retaining the Hubbard
  interaction only on a ribbon edge.  The Hubbard interaction greatly
  suppresses charge currents along the edge and promotes edge magnetism, but
  leaves the single-particle signatures of the helical liquid intact.
\end{abstract}

\pacs{03.65.Vf,71.10.Pm,71.27.+a,71.30.+h}

\maketitle

The $Z_2$ topological band insulator (TBI) \cite{HaKa10} arises from
spin-orbit (SO) coupling and is invariant under time reversal symmetry. The
bulk is insulating and the edge states, coined helical Luttinger liquids, show
gapless spin and charge excitations. An explicit realization is given by the
Kane-Mele (KM) Hamiltonian \cite{KaneMele05} which reduces to two separate
Haldane models \cite{Ha88}, with opposite signs of the Hall conductivity in
the two spin sectors. Time reversal symmetry protects the edge states against
potential scattering and weak electron-electron interactions
\cite{Wu06,Cenke06}, and allows for experimental realizations
\cite{Koenig07,HsXiQi09}.  Previous work on correlation effects has
essentially followed two routes: interaction driven topological insulators
\cite{RaQiHo08,Kai_Sun09,Yi_Zhang09,WeRuWa10,DzSuGa10}, or (as here) the
interplay of spin-orbit coupling and Coulomb repulsion
\cite{PeBa10,VaSuRi10,RaHu10,So10,Go10}.  We present the first quantum Monte
Carlo (QMC) results which document (\textit{i}) a quantum phase transition
between the quantum spin liquid (QSL) phase of \cite{Meng10} and the TBI,
(\textit{ii}) the stability of the TBI against magnetic ordering, and
(\textit{iii}) the role of fluctuations in the helical edge states of the TBI.

\begin{figure}[b]
  \includegraphics*[width=\columnwidth]{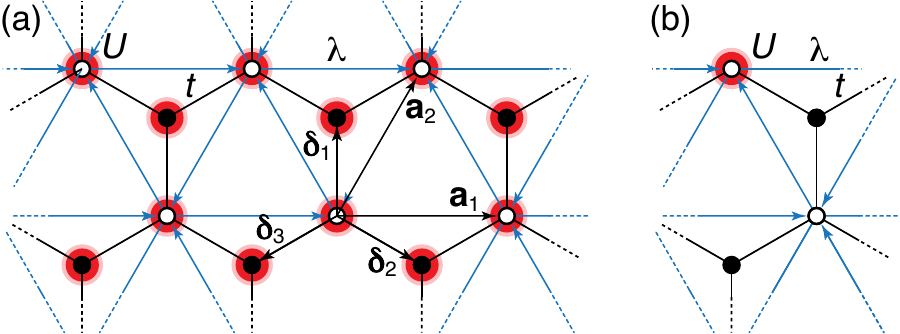}
  \caption{\label{Lattice.fig} (Color online) (a) Periodic lattice structure
    of the KM-Hubbard model with nearest-neighbor hopping $t$, spin-orbit
    coupling $\lambda$ and Coulomb repulsion $U$. Arrows indicate the current
    direction associated with the spin-orbit term for one spin species and
    sublattice.  (b) Effective model on a semi-infinite ribbon with periodic
    boundaries in the $\mathbf{a}_1$ direction, and Coulomb repulsion $U$
    only at the edge sites.}
\end{figure}

Our starting point is the KM-Hubbard model on the honeycomb lattice with
Hamiltonian ${H=H_\text{KM}+H_U}$,
\begin{align}\label{eq:KMH}
  H_{\mbox{\scriptsize{KM}}} &= -t \sum_{\langle \vec{i},\vec{j} \rangle}
  c^{\dagger}_{\vec{i}} c^\nag_{\vec{j}} + \mbox{i}\,\lambda \sum_{
    \langle\!\langle \vec{i},\vec{j} \rangle\!\rangle} c^{\dagger}_{\vec{i}}
  \vec{e}_{\vec{i},\vec{j}} \cdot {\boldsymbol\sigma} c^\nag_{\vec{j}} \,,\nonumber \\
  H_U &= \frac{U}{2} \sum_{\vec{i}} (c^{\dagger}_{\vec{i}} c^\nag_{\vec{i}} -
  1 )^2\,.
\end{align}
The spinor ${c^{\dagger}_{\vec{i}} =
  \big(c^{\dagger}_{\vec{i},\uparrow},
  c^{\dagger}_{\vec{i},\downarrow}\big)}$ creates an electron in a Wannier
state at site $\vec{i}$, ${\langle\vec{i},\vec{j}\rangle}$
means summation over the three nearest neighbors ${\vec{j} = \vec{i} +
  \vec{\boldsymbol\delta}_{n}}$ with {$\boldsymbol{\delta}_n\in\{\pm
  \boldsymbol{\delta}_1,\pm\boldsymbol{\delta}_2,\pm\boldsymbol{\delta}_3\}$},
see Fig.~\ref{Lattice.fig}(a), ${\langle\!\langle
  \vec{i},\vec{j} \rangle\!\rangle}$ denotes summation over next-nearest neighbors
${\vec{j} = \vec{i} + \vec{\boldsymbol\delta}_{n} +
  \vec{\boldsymbol\delta}_{m}}$, ${\vec{e}_{\vec{i}, \vec{j}} =
  \vec{\boldsymbol\delta}_{n} \times \vec{\boldsymbol\delta}_{m} / |
  \vec{\boldsymbol\delta}_{n} \times \vec{\boldsymbol\delta}_{m} |}$ and
${\boldsymbol\sigma}$ is the vector of Pauli matrices. 
At the particle-hole symmetric point, this model can be investigated with a
variety of QMC algorithms without encountering the
infamous negative sign problem. We present two sets of
simulations to extract bulk and boundary properties.

{\it Bulk phase diagram.}---For bulk simulations we use the projective
auxiliary field QMC approach. The ground state $| \Psi_0 \rangle $ is
filtered out of a trial wave function ${|\Psi_\text{T} \rangle}$ with
${\langle\Psi_\text{T} |\Psi_0\rangle \neq 0}$; a very good choice
  is the ground state of the KM model. For an arbitrary observable $\langle
\Psi_0 | O |\Psi_0 \rangle = \lim_{\Theta \rightarrow \infty} \langle
\Psi_\text{T} |\mbox{e}^{-\Theta H/2} O \mbox{e}^{-\Theta H/2} |
\Psi_\text{T} \rangle / \langle \Psi_\text{T} |\mbox{e}^{-\Theta H} |
\Psi_\text{T} \rangle $. The absence of the negative sign problem at half
filling follows from the fact that after a discrete Hubbard-Stratonovich
transformation of $H_U$ and subsequent integration over the fermionic degrees
of freedom, the fermionic determinants in the up and down spin sectors are
linked via complex conjugation such that their product is positive. We employ
an SU(2) invariant Hubbard-Stratonovich transformation and an imaginary time
discretization of ${\Delta\tau t = 0.1}$.  Projection parameters ${\Theta t =
  40}$ prove sufficient for converged (within statistical errors)
ground-state results.  For details of the algorithm see \cite{AsEv08}.

\begin{figure}[tp]
  \includegraphics*[width=\columnwidth]{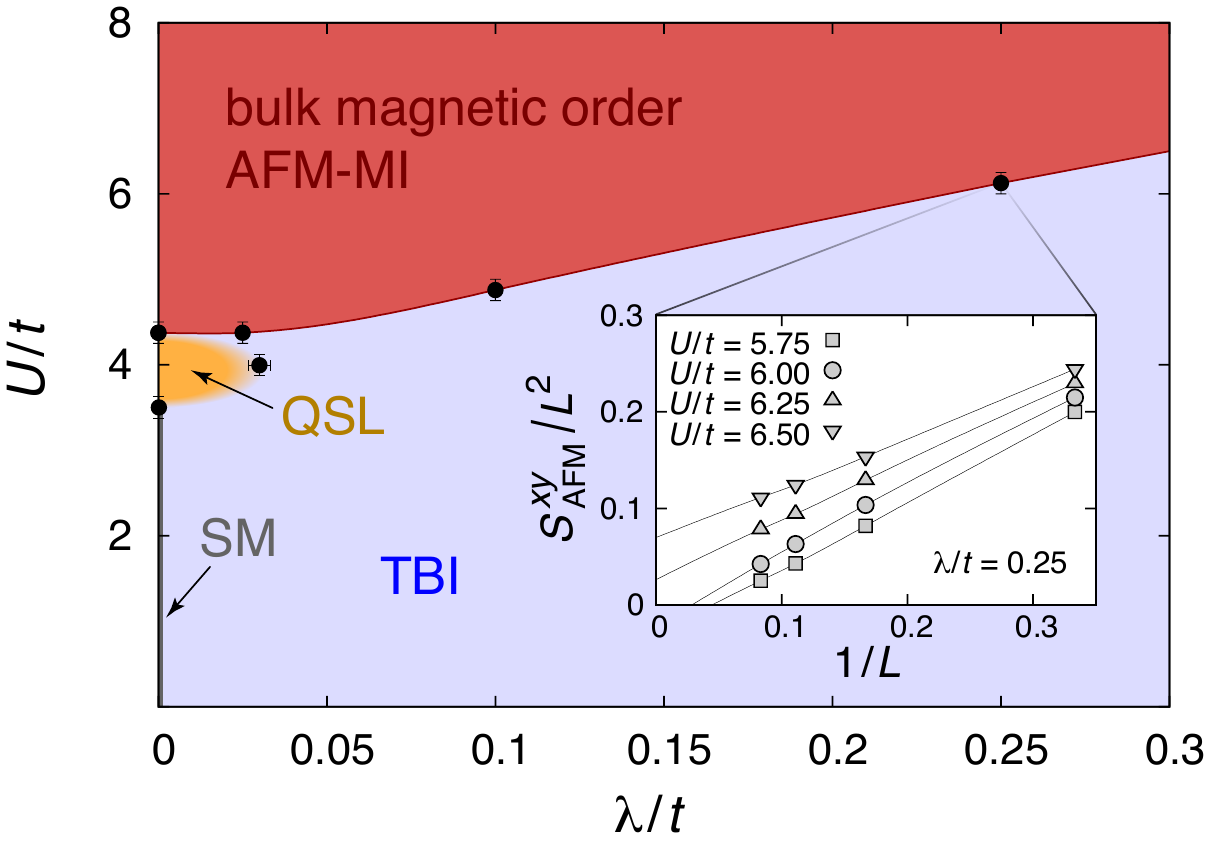}
  \caption{\label{Phasediagram.fig} (Color online) Phase diagram of
    the KM-Hubbard model from QMC simulations. Bullets correspond to computed
    phase boundaries. The four phases are the semi metal (SM), the quantum
    spin liquid (QSL), the topological or quantum spin-Hall insulator (TBI),
    and the antiferromagnetic Mott state (AFM-MI). 
    Inset: Ground-state spin-spin correlation function
    [Eq.~(\ref{SpinSpin.Eq})] for an ${L \times L}$ honeycomb lattice with
    periodic boundary conditions and ${\lambda/t=0.25}$. Lines are fits to
    the form $ a + b/L + c/L^2$. Negative values indicate that the data decay
    quicker than $1/L^2$.  Here and in subsequent figures, errorbars are
    omitted if smaller than the symbol size.}
\end{figure}

The SO coupling reduces the $SU(2)$ spin symmetry to a $U(1)$ symmetry 
corresponding to spin rotations  around the $z$-axis. The Hubbard interaction
promotes transverse, $x$-$y$ magnetic ordering \cite{RaHu10}  which can be
tracked by computing the antiferromagnetic (AFM) structure factor
\begin{equation}
  \label{SpinSpin.Eq}
  S_{\mbox{\scriptsize{AFM}}}^{xy} 
  = \frac{1}{L^2}\sum_{\vec{i},\vec{j}} (-1)^{\vec{i} + \vec{j}} \langle \Psi_0
  | S^{+}_{\vec{i}} S^{-}_{\vec{j}} + S^{-}_{\vec{i}} S^{+}_{\vec{j}} | \Psi_0 \rangle\;,
\end{equation} 
on $L\times L$ honeycomb lattices with periodic boundary conditions. At
$\lambda/t= 0.25$, this quantity is plotted versus lattice size for various
values of $U/t$ in the inset of Fig.~\ref{Phasediagram.fig}. The onset of
long-range order occurs in the region ${6 < U_\text{c}/t < 6.25}$.
Because of the underlying $U(1)$ symmetry  the quantum phase
transition between the magnetically ordered and disordered phases is expected
to be in the 3D-$XY$ universality class. Figure~\ref{Phasediagram.fig} shows
$U_\text{c}/t$ as a function of $\lambda/t$ and thus defines the magnetic phase
diagram. 
 
Several aspects of Fig.~\ref{Phasediagram.fig} deserve comments:
(\textit{i})  With the important exception of the QSL phase, the qualitative
aspects of the  magnetic phase diagram were obtained at the mean-field
level \cite{RaHu10}. The magnetic instability smoothly converges 
to the ${\lambda/t = 0}$ result \cite{Meng10}. For  
${0.025 < \lambda/t <  0.25}$ we observe no spin ordering along the $z$ quantization axis up to ${U/t=9}$.  
(\textit{ii}) The ${U/t = 0}$ line corresponds to
the KM model and describes a TBI with a single-particle gap set by $\lambda$. (\textit{iii}) The
${\lambda/t = 0}$ line has been investigated in detail in
Ref.~\cite{Meng10}. Up to ${U/t = 3.5}$ the semimetallic (SM) phase remains
stable and magnetic order sets in from ${U/t = 4.3}$ onwards. The
intermediate phase shows both spin and single-particle gaps and corresponds
to a QSL. 

\begin{figure}[t]
  \includegraphics*[width=\columnwidth]{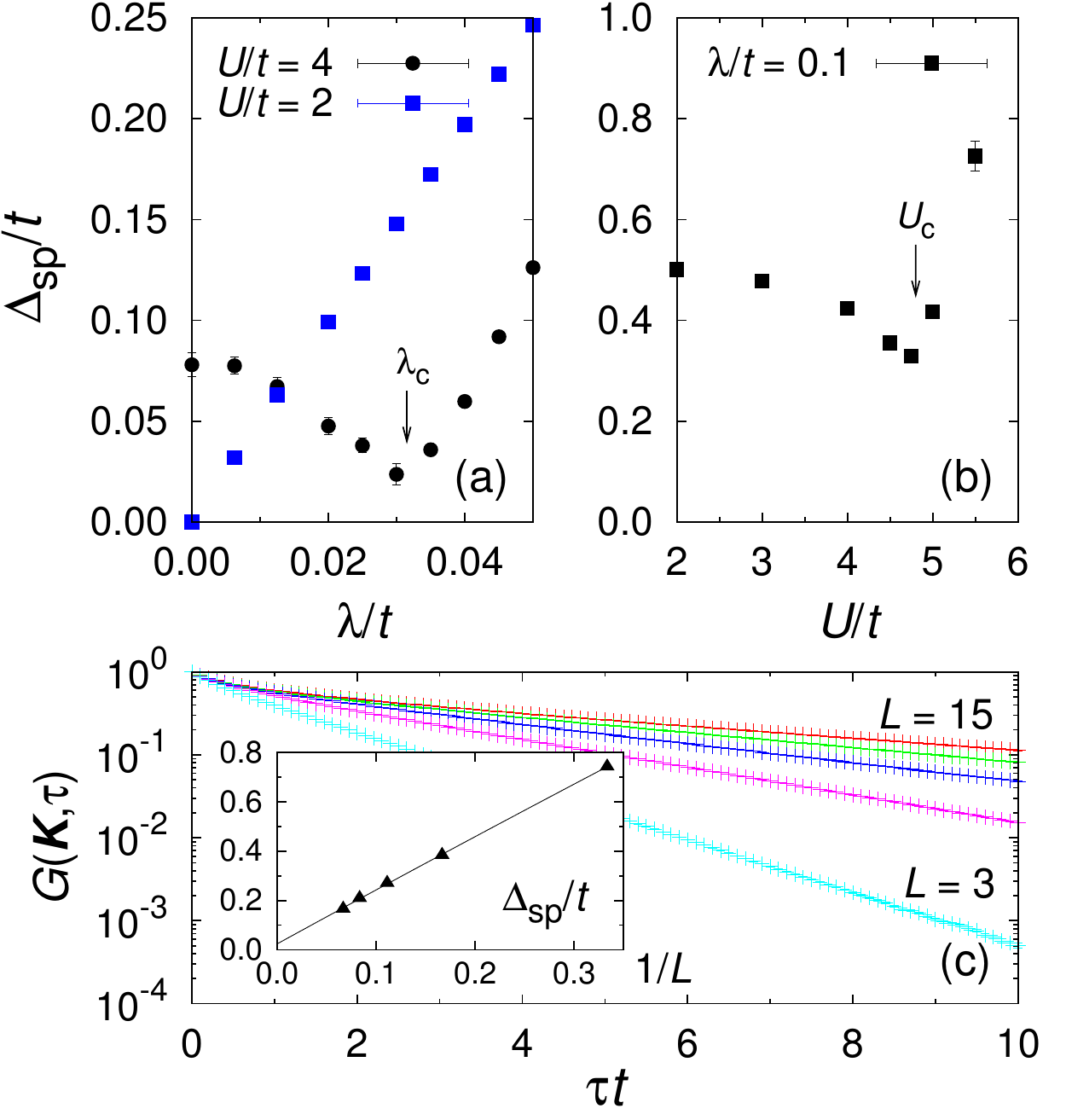}
  \caption{\label{QP.fig} (Color online) (a),(b) Single-particle gap along
    different cuts in Fig.~\ref{Phasediagram.fig}
    ($\lambda_\text{c}/t\simeq0.03$, $U_\text{c}/t\simeq4.9$).  (c) Raw data
    and size extrapolation (inset) at $U/t = 4$, $\lambda/t = 0.03$.}
\end{figure}

To further  investigate the phase diagram and in  particular the  evolution
of the QSL upon switching on the SO  coupling, we have computed the single
particle gap, $\Delta_\text{sp}$, at the Dirac point $\mathbf{K}$. This quantity is
extracted by fitting the tail of the single-particle imaginary time Green
function, $G(\mathbf{K},\tau)=\langle \Psi_0 | c^{\dagger}_{\mathbf{K},\sigma}(\tau)
c^{\nag}_{\mathbf{K},\sigma} | \Psi_0 \rangle $ [see Fig.~\ref{QP.fig}(c)],  
to the form $Z  e^{-\tau \Delta_\text{sp}} $, where $Z$ corresponds to the quasiparticle residue.
The extrapolated (in $L$) value of  $\Delta_\text{sp}$ is plotted
in Figs.~\ref{QP.fig}(a),(b) along different cuts of the phase diagram.
Starting in the SM phase, at $U/t=2$, $\Delta_\text{sp} \propto \lambda $ as
for the $U/t = 0$ case and characteristic of the TBI state \cite{KaneMele05}.
In contrast, in the QSL phase at $U/t = 4 $, $\Delta_\text{sp} $ initially 
decreases with increasing $\lambda$ but grows again for $\lambda /t \gtrsim 0.03$. We
interpret this cusp feature as a signature of a quantum phase
transition between the QSL state and the TBI  at $\lambda_c/t \simeq 0.03$.
The data support the vanishing of the  single-particle  gap at
$\lambda_c/t$ \cite{gap}. Figure~\ref{QP.fig}(b) shows that the magnetic transition as a function of
$U/t$ at fixed $\lambda/t=0.1$ is equally  apparent;  $\Delta_\text{sp} $ smoothly evolves from its
$U/t=0$ value and exhibits a cusp feature at
$U_\text{c}/t\simeq4.9$. Magnetic order breaks time reversal symmetry and
lifts the topological protection, so that $\Delta_\text{sp}$ does
not have to close at $U_\text{c}$. This evolution of the gap
can be qualitatively reproduced at the mean-field level.
From the single-particle gap, we identify four distinct phases: (\textit{i}) a  TBI
phase, where $\Delta_\text{sp}$ evolves smoothly to its $U/t=0$  value,
(\textit{ii}) a magnetically ordered MI, (\textit{iii}) a
SM line and (\textit{iv}) a QSL phase. 
\begin{figure}[t]
  \includegraphics*[width=0.95\columnwidth]{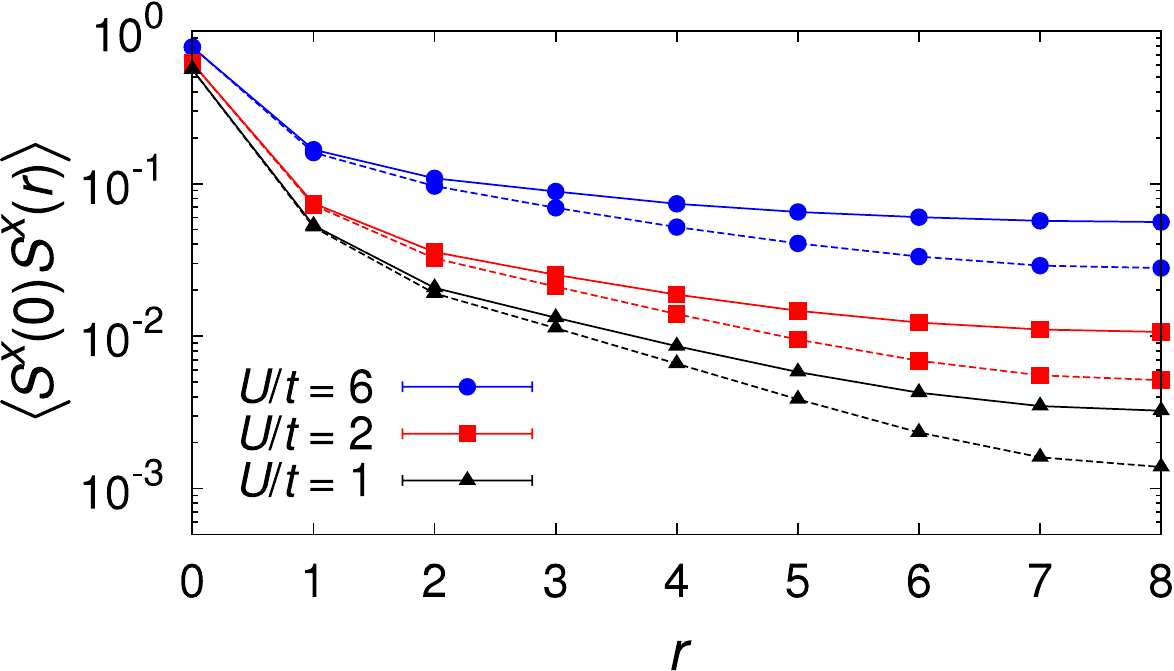}
  \caption{\label{static.fig} (Color online) Transverse spin
    correlations along the edge of the ribbon. Results are for
    ${L = 16}$, $\beta t = 20$ (dashed lines) and ${\beta t =
      40}$ (solid lines). Lines are guides to the eye.
  }
\end{figure}

{\it Edge states in the TBI phase.}--- Edge states are a hallmark feature of
TBIs.  A detailed understanding of correlation effects in these
one-dimensional liquids is crucial for theory and experiment.  To study the
helical Luttinger liquid formed at the edge of the $Z_2$ TBI, we consider the
ribbon topology of Fig.~\ref{Lattice.fig}(b).  For $U\geq U_\text{c}$, time
reversal symmetry is broken spontaneously and scattering between the left
spin-down, and right spin-up movers of the helical liquid is allowed, thus
opening a gap in the edge states and destroying the TBI state.  As argued
above, at $U < U_\text{c}$ the bulk is adiabatically linked to the $U/t = 0$
line. Since furthermore the helical liquid is exponentially localized on the
boundary (as readily seen in the KM model), we retain the Hubbard interaction
only on one zig-zag edge of the ribbon [cf. Fig.~\ref{Lattice.fig}(b)]. With
this ansatz, the bulk plays the role of a fermionic bath which can be
integrated out at the expense of a Gaussian integral. This yields an
effective one-dimensional action,
%
\begin{align}
  \label{Action.Eq}
  \mathcal{S} = & -\sum_{\sigma,r,r'} \int_{0}^{\beta} {\rm d} \tau
  \int_{0}^{\beta} {\rm d} \tau' 
  c^{\dagger}_{r,\sigma} G_{0,\sigma}^{-1} (r-r') c^\nag_{r',\sigma} \nonumber \\
  & + U \sum_{r} \int_{0}^{\beta} 
  \left[n_{r,\uparrow}(\tau) -\oh\right] \left[n_{r,\downarrow}(\tau) - \oh\right]\,,
\end{align}
where $r$ is an edge site index, and ${G_{0,\sigma}(r-r')}$ is the free Green
function of the KM model on the ribbon topology. We can solve the
action~(\ref{Action.Eq}) exactly using the weak-coupling expansion
continuous-time QMC method \cite{Rubtsov05,Luitz09}, on arbitrarily wide
$L\times L'$ ribbons (here $L'=64$).  The validity of this effective model at
$U<U_\text{c}$ has been verified by QMC calculations for the full
model~(\ref{eq:KMH}) on narrow ribbons.  We take $\lambda/t=0.25$ in the
following.

\begin{figure}[t]
  \includegraphics*[width=\columnwidth]{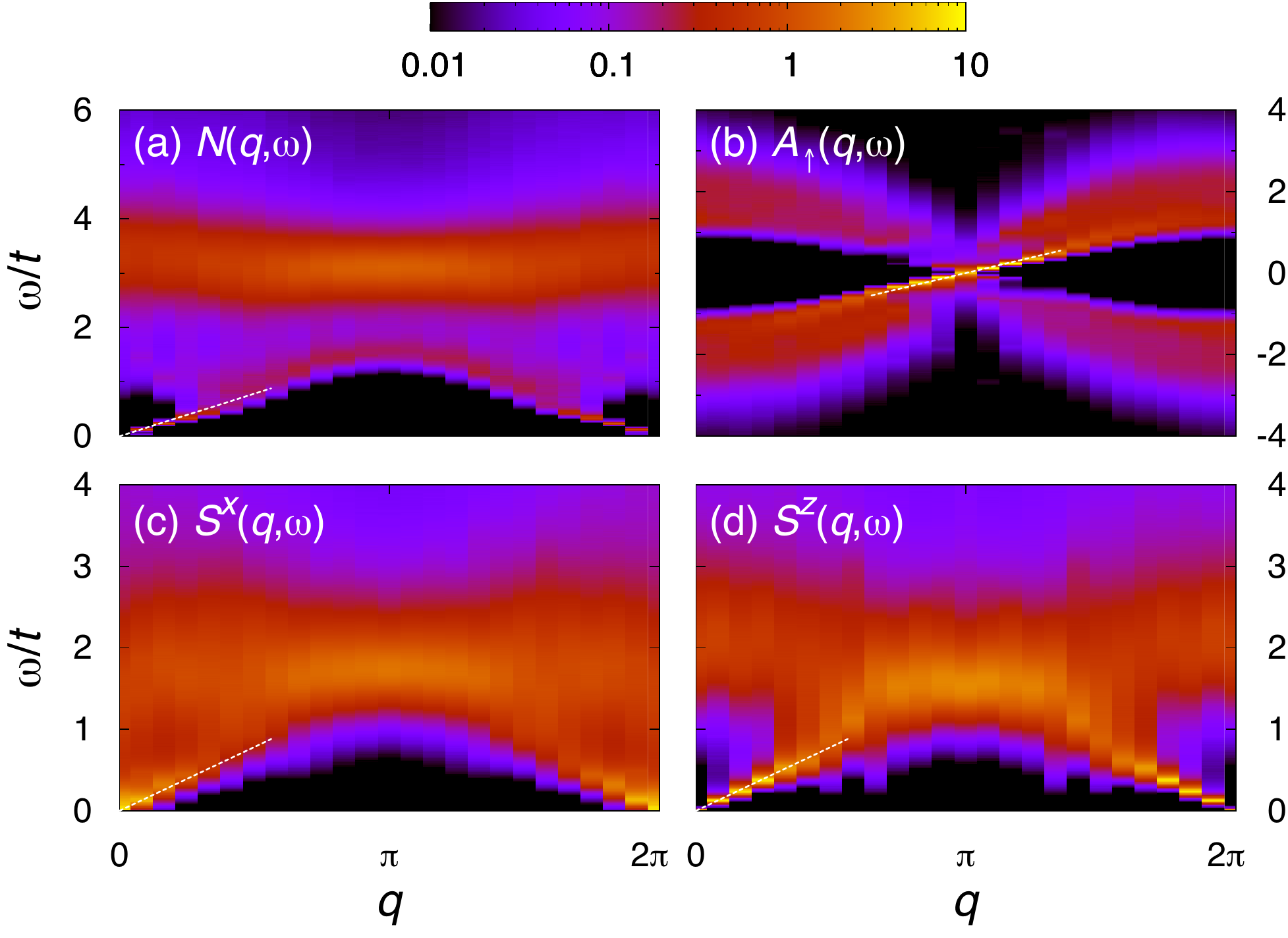}
  \caption{\label{DynU2.fig} (Color online) Dynamic spectral functions in
    the (a) charge sector, (b) spin-resolved one-particle sector, and (c,d)
    spin sector, measured along the edge. The parameters are $U/t=2$,
    $L=24$, $L'=64$, $\lambda/t=0.25$ and $\beta t=40$.
    Dotted lines show the velocities of the free helical liquid ($U/t=0$).}
\end{figure}

At $U/t=0$ and half filling, the dispersion relation of the helical liquid
satisfies ${\varepsilon_{q,\uparrow} = - \varepsilon_{q,\downarrow}}$, and
the edge states are unstable towards transverse {\it ferromagnetic order}.
Figure~\ref{static.fig} shows the development of substantial spin-spin
correlations in the transverse direction with decreasing temperature and
increasing $U/t$.  This corresponds to the dominant correlation function.

We calculate dynamic structure factors along the edge,
\begin{equation}
  \label{Dyn_ph.eq}
  O(q,\omega) = \frac{1}{Z} \sum_{n,m} \mbox{e}^{-\beta E_n} 
  |\langle m | O(q) | n \rangle|^{2} \delta( E_m - E_0 - \omega).
\end{equation} 
For charge, $N(q,\omega)$, $O(q)=N(q) =\frac{1}{\sqrt{L}}\sum_{r}
\mbox{e}^{\mbox{\scriptsize{i}} q r } n_r$, and for spin,
$S^{\alpha}(q,\omega)$, $O(q)=S^{\alpha}(q) = \frac{1}{\sqrt{L}}\sum_{r}
\mbox{e}^{\mbox{\scriptsize{i}} q r } S^{\alpha}_r$. Single-particle dynamics
are deduced from the single-particle Green function via the spectral
functions $A_{\sigma}(q,\omega)=-\pi^{-1}\text{Im}\,G_{\sigma}(q,\omega)$,
where by time reversal symmetry we have ${A_{\uparrow}(q,\omega) =
  A_{\downarrow}(-q,\omega)}$.

Figure~\ref{DynU2.fig} shows these dynamic quantities at ${U/t=2}$. The
dominant features of the single-particle spectral function, see
Fig.~\ref{DynU2.fig}(b), follow the noninteracting system: within the bulk
band gap, gapless single-particle excitations emerge with a velocity tied to
the $z$-component of the spin. For $U/t=0$, the particle-hole spectra can be
deduced from the single-particle dynamics by computing the bubble. Within
this framework the dynamic charge structure factor as well as the
$z$-component of the dynamic spin structure factor are identical. Both
quantities conserve the $z$-component of spin, such that at low energies
(i.e. below the bulk gap) only particle-hole excitations within the left, or
right movers are allowed. This produces a linear mode around ${q = 0}$ as
observed in Figs.~\ref{DynU2.fig}(a),(d). At higher energies, particle-hole
excitations involving bulk states become apparent. Upon inspection of
Figs.~\ref{DynU2.fig}(a),(d) one sees that the support of both quantities is
very similar. However, the spectral weight of the low-lying charge modes is
greatly suppressed in comparison to the longitudinal spin mode. The
transverse spin susceptibility involves a spin-flip process and hence
excitations between the left and right dispersion relations. This produces a
continuum of excitations in the long-wavelength limit
[cf. Fig.~\ref{DynU2.fig}(c)].

\begin{figure}[t]
  \includegraphics*[width=\columnwidth]{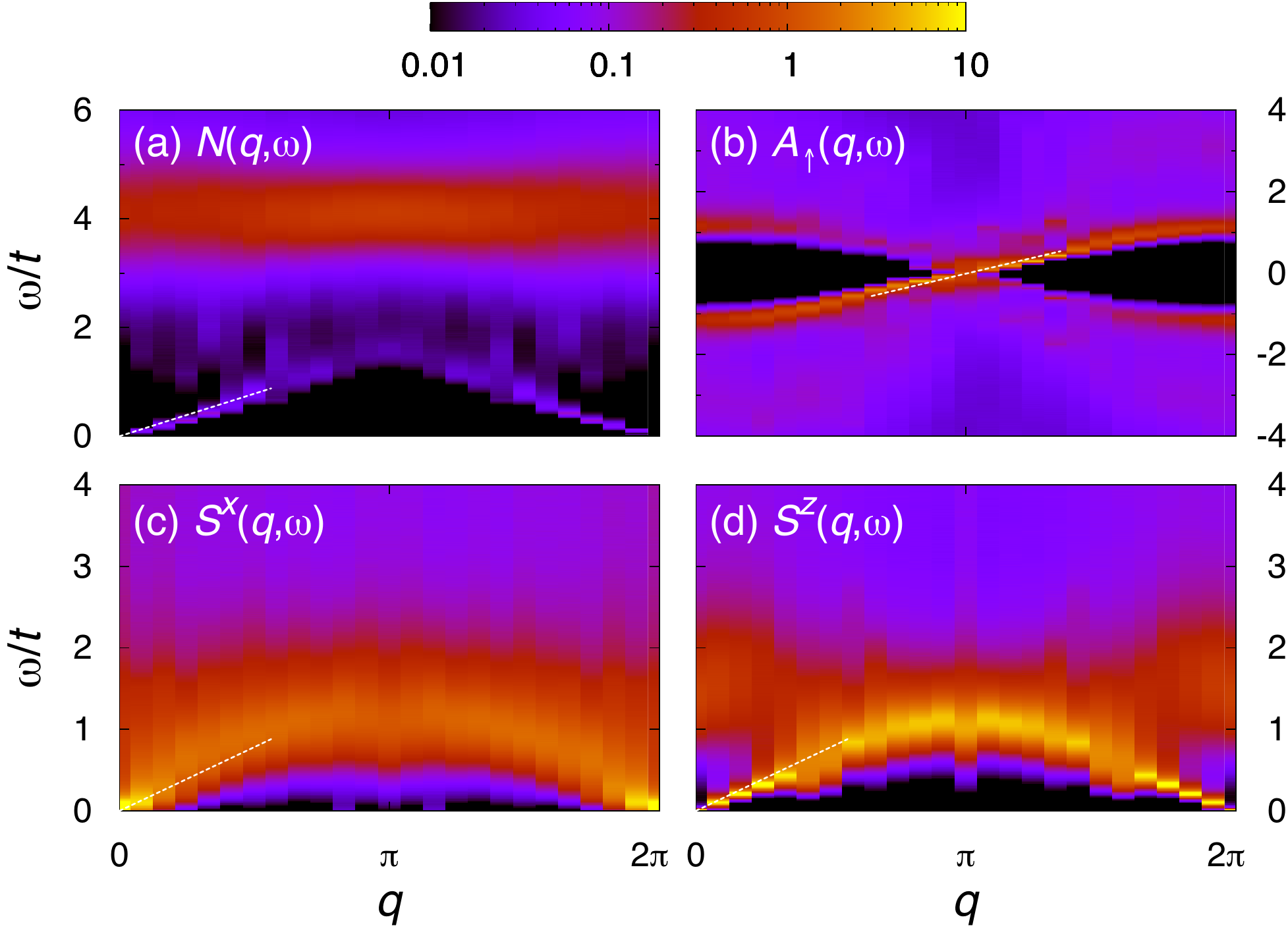}
  \caption{\label{DynU6.fig} (Color online) Same as in Fig.~\ref{DynU2.fig}
    but for ${U/t=5}$.}
\end{figure}

At large $U/t=5$ (Fig.~\ref{DynU6.fig}) we observe a strong depletion of
spectral weight in the low-lying charge modes [Fig.~\ref{DynU6.fig}(a)],
which leads to reduction of the Drude weight by 1 order of magnitude.
In contrast, despite strong correlations, the single-particle spectrum [Fig.~\ref{DynU6.fig}(b)]
still exhibits the typical signature of the
helical edge state. The growth of the equal-time transverse ferromagnetic
correlations as a function of $U/t$ (Fig.~\ref{static.fig}) leads to a
piling up of low-lying spectral weight in $S^{x}(q,\omega)$  for $q\to 0$.   

In the TBI phase, where the effective model of Eq.~(\ref{Action.Eq}) is
valid, one can argue that the charge, longitudinal
spin and Fermi velocities should be rather insensitive to the value of $U/t$ since
they are inherited from the bulk. This is confirmed
by our numerical results. Correlation effects become manifest
in a very strong variation of matrix elements in the dynamic quantities of
Eq.~(\ref{Dyn_ph.eq}). In particular, the depletion of low-lying spectral
weight in the charge sector suppresses charge transport along the edge.
In contrast, spin fluctuations as well as the signatures of the 
helical liquid in the single-particle spectra persist. 

{\it Summary.}---We have derived the bulk phase diagram of the Kane-Mele
Hubbard model from QMC simulations. We established
the exact location of the previously predicted magnetic transition at large
$U/t$ \cite{RaHu10}, and that the nonmagnetic region is dominated by the
TBI. The single-particle gap provides strong evidence for a quantum phase
transition at finite SO coupling between the QSL and the TBI. Neither of
these states can be characterized by a local order parameter, and a detailed
understanding of the transition represents a fascinating open issue.
Applying QMC to an effective model of the helical edge states,
we have studied the impact of electronic correlations by
calculating one and two-particle dynamics in the TBI phase. 
Correlation effects lead to an order of magnitude reduction
of low-lying long wavelength charge fluctuations, and thereby charge
transport, and promote  transverse magnetic  fluctuations.   The
single-particle spectrum retains its weak-coupling features. 

 We particularly thank Z.~Y. Meng, A. Muramatsu and S. Wessel
for helpful discussions. We acknowledge conversations with C. Xu, M. Imada
and Y. Yamaji, and support from DFG Grants No.~AS120/4-3 (TCL) and No.~FOR1162
(MH) and NSF No.~PHY05-51164 (TCL, FFA).
We thank the LRZ Munich and the J\"{u}lich Supercomputing
Centre for generous allocation of CPU time.

{\it Note added.}---{After completion of this work (arxiv: 1011.5063), a QMC investigation
\cite{Zheng10} and two approximate studies \cite{Yamaji10,Yu10} of the same model came out.


\begin{thebibliography}{10}

\bibitem{HaKa10}
M.~Z. Hasan and C.~L. Kane, Rev. Mod. Phys. {\bf 82},  3045  (2010).

\bibitem{KaneMele05}
C.~L. Kane and E.~J. Mele, Phys. Rev. Lett. {\bf 95},  226801  (2005).

\bibitem{Ha88}
F.~D.~M. Haldane, Phys. Rev. Lett. {\bf 61},  2015  (1988).

\bibitem{Wu06}
C. Wu, B.~A. Bernevig, and S.-C. Zhang, Phys. Rev. Lett. {\bf 96},  106401
  (2006).

\bibitem{Cenke06}
C. Xu and J.~E. Moore, Phys. Rev. B {\bf 73},  045322  (2006).

\bibitem{Koenig07}
M. K\"onig {\it et al.}, Science {\bf 318},  766  (2007).

\bibitem{HsXiQi09}
D. Hsieh {\it et al.}, Nature {\bf 460},  1101
  (2009).

\bibitem{RaQiHo08}
S. Raghu, X. Qi, C. Honerkamp, and S. Zhang, Phys. Rev. Lett. {\bf 100},
  156401  (2008).

\bibitem{Kai_Sun09}
K. Sun, H. Yao, E. Fradkin, and S.~A. Kivelson, Phys. Rev. Lett. {\bf 103},
  046811  (2009).

\bibitem{Yi_Zhang09}
Y. Zhang, Y. Ran, and A. Vishwanath, Phys. Rev. B {\bf 79},  245331  (2009).

\bibitem{WeRuWa10}
J. Wen, A. R{\"u}egg, C.~J. Wang, and G.~A. Fiete, Phys. Rev. B {\bf 82},
  075125  (2010).

\bibitem{DzSuGa10}
M. Dzero, K. Sun, V. Galitski, and P. Coleman, Phys. Rev. Lett. {\bf 104},
  106408  (2010).

\bibitem{PeBa10}
D. Pesin and L. Balents, Nat. Phys. {\bf 6},  376  (2010).

\bibitem{VaSuRi10}
C.~N. Varney, K. Sun, M. Rigol, and V. Galitski, Phys. Rev. B {\bf 82},  115125
   (2010).

\bibitem{RaHu10}
S. Rachel and K.~Le Hur, Phys. Rev. B {\bf 82},  075106  (2010).

\bibitem{So10} 
D. Soriano and J. Fern\'{a}ndez-Rossier, Phys. Rev. B {\bf 82}, 161302 (2010).

\bibitem{Go10}
J. Goryo and N. Maeda, arXiv:1007.4671v2.

\bibitem{Meng10}
Z.~Y. Meng {\it et al.}, Nature {\bf
  464},  847  (2010).

\bibitem{AsEv08}
F.~F. Assaad and H.~G. Evertz, Lect. Notes Phys. {\bf 739},  277  (2008).

\bibitem{gap}
Transitions  between trivial  and topological band insulators can  only occur  via a metallic state.

\bibitem{Rubtsov05}
A.~N. Rubtsov, V.~V. Savkin, and A.~I. Lichtenstein, Phys. Rev. B {\bf 72},
  035122  (2005).

\bibitem{Luitz09}
D. Luitz and F.~F. Assaad, Phys. Rev. B {\bf 81},  024509  (2010);
F.~F. Assaad and T.~C. Lang, Phys. Rev. B {\bf 76},  035116  (2007).

\bibitem{Zheng10}
D. Zheng, C. Wu and G.-M. Zhang, arXiv:1011.5858v1.

\bibitem{Yamaji10}
Y. Yamaji and M. Imada, arXiv:1012.2637v1.

\bibitem{Yu10} 
S.-L. Yu, X.~C. Xie and J.-X. Li, arXiv:1101.0911v1.

\end{thebibliography}
\end{document}